\documentclass[12p]{article}

\begin{document}

\input amssym.tex

\title{The quantum theory of scalar fields on the de Sitter expanding universe}

\author{Ion I.  Cot\u aescu\thanks{E-mail:~~~cota@physics.uvt.ro} ,
Cosmin Crucean\thanks{E-mail:~~~} and Adrian Pop\thanks{E-mail:~~~}\\
{\small \it West University of Timi\c soara,}\\
{\small \it V.  P\^ arvan Ave.  4, RO-300223 Timi\c soara, Romania}}

\maketitle

\begin{abstract}
New quantum modes of the free scalar field are derived in a special
time-evolution picture that may be introduced in moving charts of de Sitter
backgrounds. The wave functions of these new modes are solutions of the
Klein-Gordon equation and energy eigenfunctions, defining the energy basis.
This completes the scalar quantum mechanics where the momentum basis is
well-known from long time. In this enlarged framework the quantization of the
scalar field can be done in canonical way obtaining the principal conserved
one-particle operators and the Green functions.

 Pacs:
04.62.+v
\end{abstract}

\newpage

\section{Introduction}

The quantum scalar field is the main piece in problems concerning quantum
effects in the presence of gravitation or in quantum gravity and cosmology. Of
a special interest in cosmology is the de Sitter (dS) expanding universe
carrying scalar fields variously coupled with gravitation. In the quantum
theory a special role is played by the free fields (minimally coupled with
gravitation) since these are the principal ingredients in calculating
scattering amplitudes using perturbations.

In curved spacetimes the field equations have covariant forms such that their
operators are globally defined. If the manifold has, in addition, some
isometries then the corresponding Killing vectors give rise to the generators
of the representations of the isometry group carried by spaces of mater fields
\cite{ES}. These generators are operators of the quantum mechanics whose main
virtue is to commute with the operator of the field equation. On the other
hand, these operators are related to the conserved quantities predicted by the
Noether theorem from which one constructs the conserved one-particle operators
of the quantum field theory. For this reason we say here that the isometry
generators, or any other operator which commutes with that of the field
equation, are {\em conserved} operators of the relativistic quantum mechanics.
In general, the quantum modes can be globally defined on curved manifolds using
field equations and suitable sets of commuting conserved operators. This method
is useful in the case of the four-dimensional dS background which has $SO(4,1)$
isometries.

The quantum  modes of the scalar field in moving frames of dS manifolds are
well-known from long time \cite{BD,WALD}. Their wave functions are solutions of
the Klein-Gordon (KG) equation and eigenfunctions of the conserved momentum
operator but these are not able to provide us with information about energy.
This is because in dS moving frames the operator $i\partial_t$ is not a Killing
vector and, therefore, it is not conserved. We suggested \cite{CD0,CD1} that
the correct energy operator must be the time-like Killing vector field of the
dS geometry, as in the case of the anti-de Sitter manifolds \cite{AIS,CADS}.
Moreover, we studied other conserved operators arising from  dS isometries,
showing that the energy operator does not commute with the momentum one
\cite{CD1,CSP}. This means that the momentum and energy can not be measured
simultaneously with desired accuracy. In other respects, it is known that there
are no mass-shells. Obviously, these major difficulties may be avoided using
new special methods for analyzing the time evolution of the free fields.

Recently we constructed a new Dirac quantum mechanics on spatially flat
Robertson-Walker spacetimes in which we defined different time evolution
pictures \cite{CSP}. In the case of dS backgrounds these pictures helped us to
find new solutions of the Dirac equation and to understand how can be measured
the momentum and energy. We started with the {\em natural} picture (NP) which
is just the usual Dirac theory in dS moving frames as it results from its
Lagrangian. In this picture we found the complete system of fundamental spinors
with well-determined momentum and helicity \cite{CD1}. The NP can be
transformed in a new time-evolution picture where the Dirac equation do not
depend explicitly on time. This is the Schr\" odinger picture (SP) we have
introduced for studying the energy quantum modes of the Dirac field which can
not be derived in other conjectures \cite{CSP}. In  SP  we derived the complete
set of fundamental solutions of the Dirac equation of given energy, momentum
direction and helicity \cite{CD2}. We must specify that our SP defined at the
level of the relativistic quantum mechanics is completely different from the
{\em functional} Schr\" odinger-picture of the quantum field theory \cite{BOG}
which was used for studying scalar fields \cite{GLH}.

Here we  would like to continue our study using the same approach for improving
the quantum theory of the massive charged scalar field in dS moving frames. We
start with the NP where the time evolution is governed by the well-known KG
equation which depends explicitly on time \cite{BD}. This produces quantum
modes depending on momentum whose wave functions constitute a complete set of
normalized fundamental solutions, defining the momentum basis. As in the Dirac
case, we can define the SP where the KG equation does not depend explicitly on
time. In this picture we derive the new complete set of normalized fundamental
solutions of given energy and momentum direction which define the energy basis.
Using these bases a coherent quantum mechanics can be constructed as a starting
point to a suitable quantum theory of fields. Our objective here is to follow
this way deriving step by step all the elements of the quantum theory of the
scalar field in dS moving frames, up to the form of the Green functions related
to the commutator ones.

We start in the second section with a short introduction in the scalar quantum
mechanics on dS backgrounds defining the new SP. The next section is devoted to
the fundamental solutions of the momentum and, respectively, energy bases. The
wave functions of the energy basis are studied giving details since these
represent the main new result obtained here. The transition coefficients
between these bases are also written down in NP. In the fourth section we
perform the quantization in canonical manner obtaining the form of the
conserved one-particle operators in both the bases we use. Finally, we present
the the properties of the principal Green functions related to the commutator
ones. The concluding remarks includes some observations on some specific
features of the energy and momentum measurements on dS spacetimes.

\section{Scalar quantum mechanics on dS spacetimes}

Let us consider $M$ be the dS spacetime, defined as the hyperboloid of radius
$\frac{1}{\omega}$ in the five-dimensional flat spacetime $M^5$ of coordinates
$z^A$ $(A,\,B,...= 0,1,2,3,5)$ and metric $\eta^5={\rm diag}(1,-1,-1,-1,-1)$
\cite{SW}. The hyperboloid equation $\omega^2\eta^5_{AB}z^A z^B=-1$  defines
$M$ as the homogeneous space of the pseudo-orthogonal group $SO(4,1)$ which is
at the same time the gauge group of the metric $\eta^{5}$ and the isometry
group, $I(M)\equiv SO(4,1)$, of the dS spacetime. Then, it is convenient to use
the covariant  real parameters $\xi^{AB}=-\xi^{BA}$ since in this case the
orbital basis-generators of the scalar representation of $SO(4,1)$, carried by
the space of the scalar fields over $M^{5}$, have the standard form
\begin{equation}\label{LAB5}
 L_{AB}^{5}=i\left[\eta_{AC}^5 z^{C}\partial_{B}-
 \eta_{BC}^5 z^{C}
\partial_{A}\right].
\end{equation}
They  will give us directly the Killing vectors $k_{(AB)}$ of $M$ which define
the basis-generators
\begin{equation}
L_{(AB)}=-ik^{\mu}_{(AB)}\partial_{\mu}
\end{equation}
of the scalar representation of  $I(M)$. These  operators can be calculated in
any local chart $\{x \}$ of coordinates $x^{\mu}$ ($\mu,\nu,...=0,1,2,3$) where
we know the functions $z^A(x)$.

In an arbitrary chart $\{ x\}$  the action of a charged scalar field $\phi$ of
mass $m$, minimally coupled with the gravitational field, reads
\begin{equation}\label{action}
{\cal S}[\phi,\phi^*]=\int d^4 x \sqrt{g}\,{\cal L}=\int d^4 x
\sqrt{g}\left(\partial^{\mu}\phi^*\partial_{\mu}\phi-m^2 \phi^* \phi\right)\,,
\end{equation}
where $g=|\det(g_{\mu\nu})|$. This action gives rise to the KG equation
\begin{equation}\label{KG}
\frac{1}{\sqrt{g}}\,\partial_{\mu}\left[\sqrt{g}\,
g^{\mu\nu}\partial_{\nu}\phi\right]+m^2\phi=0\,.
\end{equation}
The conserved quantities  predicted by the Noether theorem can be calculated
with the help of the stress-energy tensor
\begin{equation}
T_{\mu\nu}=\partial_{\mu}\phi^*\partial_{\nu}\phi+
\partial_{\nu}\phi^*\partial_{\mu}\phi-g_{\mu\nu}{\cal L}\,.
\end{equation}
Thus, for each isometry corresponding to a Killing vector $k_{(AB)}$ there
exists the conserved current $\Theta^{\mu}[k_{(AB)}]=-T^{\mu \,\cdot}_{\cdot
\,\nu}k^{\nu}_{(AB)}$ which satisfies $\Theta^{\mu}[k_{(AB)}]_{;\mu}=0$
producing the conserved quantity \cite{SW}
\begin{equation}
C[k_{(AB)}]=\int_{\Sigma} d\sigma_{\mu} \sqrt{g}\,\, \Theta^{\mu}[k_{(AB)}]\,,
\end{equation}
on a given hypersurface $\Sigma \subset M$. Moreover, generalizing the form of
the conserved electric charge due to the internal $U(1)$ symmetry one defines
the relativistic scalar product of two scalar fields as \cite{BD}
\begin{equation}\label{SP1}
\langle \phi,\phi'\rangle=i\int_{\Sigma} d\sigma^{\mu}\sqrt{g}\, \phi^*
\stackrel{\leftrightarrow}{\partial_{\mu}} \phi'\,,
\end{equation}
using the notation $f\stackrel{\leftrightarrow}{\partial}h=f(\partial
h)-h(\partial f)$. With this definition one obtains the following identities
\begin{equation}\label{cLAB}
C[k_{(AB)}]=\langle \phi,L_{(AB)}\phi\rangle
\end{equation}
which can be proved for any Killing vector using the field equation (\ref{KG})
and the Green's theorem. These identities will be useful in quantization,
giving directly the conserved one-particle operators of the quantum field
theory.

As in the Dirac case \cite{CSP}, we say that the NP is the genuine quantum
theory in the chart $\{t,{\bf x}\}$ with Cartesian coordinates and the
Robertson-Walker line element
\begin{equation}\label{mrw}
ds^2=g_{\mu\nu}(x)dx^{\mu}dx^{\nu}=dt^2-e^{2\omega t} (d{\bf x}\cdot d{\bf
x})\,,
\end{equation}
where $\sqrt{g}=e^{3\omega t}$. In this picture the time evolution of the
massive scalar field is governed by the KG equation,
\begin{equation}\label{KG1}
\left( \partial_t^2-e^{-2\omega t}\Delta +3\omega
\partial_t+m^2\right)\phi(x)=0\,.
\end{equation}
The solutions of this equation may be square integrable functions or tempered
distributions with respect to the scalar product (\ref{SP1}) that in NP and for
$\Sigma={\Bbb R}^3$ takes the form
\begin{equation}\label{SP2}
\langle \phi,\phi'\rangle=i\int d^3x\, e^{3\omega t}\, \phi^*(x)
\stackrel{\leftrightarrow}{\partial_{t}} \phi'(x)\,.
\end{equation}

The principal operators of NP, the energy $\hat H$, momentum $\bf{\hat P}$ and
coordinate $\bf{\hat X}$, have the same forms as in special relativity,
\begin{equation}\label{ON}
(\hat H \phi)(x)=i\partial_t\phi(x)\,,\quad (\hat P^i
\phi)(x)=-i\partial_i\phi(x)\,,\quad (\hat X^i \phi)(x)=x^i\phi(x)\,.
\end{equation}
The operators $\hat X^i$ and $\hat P^i$ are time-independent and satisfy the
well-known canonical commutation relations
\begin{equation}\label{com}
[\hat X^i, \hat P^j]=i\delta_{ij}I\,,\quad [\hat H, \hat X^i]=[\hat H,\hat
P^i]=0\,,
\end{equation}
where $I$ is the identity operator.

Since the scalar theory on $M$ has the high symmetry induced by the isometry
group $I(M)$, the basis generators $L_{(AB)}$ are conserved operators commuting
with the KG operator of Eq. (\ref{KG1}). In this conjecture, we define the
conserved energy operator \cite{CD1,CSP},
\begin{equation}\label{HH}
H\equiv\omega {L}_{(05)}=\hat H+ \omega \hat{\bf X}\cdot \hat{\bf P}\,,
\end{equation}
and verify that the momentum and angular momentum  operators are also conserved
since we have
\begin{equation}\label{LLL1}
\hat P^{i} \equiv \omega\left(L_{(5i)}-L_{(0i)}\right),\quad L_{k}\equiv
\textstyle{\frac{1}{2}}\varepsilon_{ijk}{L}_{(ij)}=\varepsilon_{ijk}\hat X^i
\hat P^j\,.
\end{equation}
In addition, there exists more three conserved generators,
\begin{equation}\label{LLL2}
R^i(t)\equiv L_{(5i)}+L_{(0i)}= \left(\frac{e^{-2\omega t}}{\omega}\, I- \omega
\hat{\bf X}^2\right)\hat P^i + 2 \hat X^i H\,,
\end{equation}
which do not have an immediate physical significance \cite{CD1}. The $SO(4,1)$
transformations corresponding to these basis-generators and the associated
isometries are briefly presented in Ref. \cite{CD1}. The conserved energy
operator satisfy the commutation relations
\begin{equation}\label{HPX}
[H,\hat P^i]=i\omega \hat P^i\,, \quad [H,\hat X^i]=-i\omega \hat X^i\,,
\end{equation}
showing that the measurements of these observables are affected by uncertainty,

The NP can be changed using point-dependent operators which could be even
non-unitary operators since the relativistic scalar product does not have a
direct physical meaning as that of the non-relativistic quantum mechanics. We
exploited this opportunity for defining the Schr\" odinger picture  with the
help of the transformation $\phi(x)\to \phi_S(x)=U(x)\phi(x)$  produced by the
operator of time dependent {\em dilatations} \cite{CSP}
\begin{equation}\label{U}
U(x)=\exp\left[-\omega t(x^i \partial_i)\right]\,.
\end{equation}
This has the remarkable property
\begin{equation}\label{Udag}
U^{+}(x)= e^{3\omega t}\, U^{-1}(x) \,,
\end{equation}
and the following convenient action
\begin{equation}\label{propU}
U(x)F({x}^i)U^{-1}(x)=F\left(e^{-\omega t}{x}^i\right)\,,\quad
U(x)G(\partial_i)U^{-1}(x)=G\left(e^{\omega t}\partial_i\right)\,,
\end{equation}
upon any analytical functions $F$ and $G$. This transformation leads to the KG
equation of the SP
\begin{equation}\label{KG2}
\left[\left( \partial_t+\omega x^i\partial_i \right)^2-\Delta +3\omega
(\partial_t+\omega x^i\partial_i)+m^2\right]\phi_S(x)=0\,,
\end{equation}
and allow us to define the scalar product of this picture,
\begin{equation}\label{spSP}
\langle \phi_S,\phi_S'\rangle\equiv\langle \phi,\phi'\rangle=i\int d^3x
\left[\phi^*_S \stackrel{\leftrightarrow}{\partial_{t}} \phi_S' +\omega
x^i(\phi^*_S \stackrel{\leftrightarrow}{\partial_{i}} \phi_S')\right] \,,
\end{equation}
as it results from Eqs. (\ref{U}) and (\ref{Udag}).

The specific operators of SP, $H_S$,  $P^i_S$ and $X^i_S$,  are defined as
\begin{equation}\label{OS}
(H_S \phi_S)(x)=i\partial_t\phi_S(x)\,,\quad (P^i_S
\phi_S)(x)=-i\partial_i\phi_S(x)\,,\quad (X^i_S \phi_S)(x)=x^i\phi_S(x)\,,
\end{equation}
obeying commutation relations similar to Eqs. (\ref{com}). The meaning of these
operators can be understood in the NP.  Performing the inverse transformation
we find that
\begin{equation}\label{Ht}
U^{-1}(x)\,H_S\,U(x)= H\,,
\end{equation}
and the new  interesting time-dependent operators of NP,
\begin{eqnarray}
X^i(t)&=&U^{-1}(x)\,X_S^i\,U(x)=e^{\omega t} \hat X^i\,,\label{Xt}\\
P^i(t)&=&U^{-1}(x)\,P_S^i\,U(x)=e^{-\omega t}\hat P^i\,,\label{Pt}
\end{eqnarray}
satisfying the usual commutation relations (\ref{com}). The angular momentum
has the same expression in both these pictures since it commutes with $U(x)$.
We note that even if  $X^i(t)$ and $P^i(t)$ commute with $H$ they are not
conserved operators since they do not commute with the KG operator.

In NP picture the eigenvalues problem $Hf_E(t,{\bf x})=Ef_E(t,{\bf x})$ of the
energy operator (\ref{Ht}) leads to energy eigenfunctions of the form
\begin{equation}
f_E(t,{\bf x})=F[e^{\omega t}{\bf x}]e^{-iEt}
\end{equation}
where $F$ is an arbitrary function. This explains why in this picture one can
not find energy eigefunctions separating  variables. However, in our SP these
eigenfunctions become the new functions
\begin{equation}
f^S_E(t,{\bf x})=U(x)f_E(t,{\bf x})=F({\bf x})e^{-iEt}
\end{equation}
which have separated variables. This means that in  SP new quantum modes could
be derived using the method of separating variables in coordinates or even in
momentum representation.

\section{Scalar plane waves}

The specific feature of the quantum mechanics on $M$ is that the energy
operator (\ref{HH}) obeys Eqs. (\ref{HPX}) which means that the conserved
energy and momentum can not be measured simultaneously with desired accuracy.
Consequently, there are no particular solutions of the KG equation with
well-determined energy and momentum, being forced to consider different plane
waves solutions depending either on momentum or on energy and momentum
direction. Thus we shall work with two bases of fundamental solutions namely
the momentum and energy ones.

\subsection{The momentum basis}

It is known that the KG equation (\ref{KG1}) of NP can be analytically solved
in terms of Bessel functions \cite{BD}. There are fundamental solutions
determined as eigenfunctions of the set of commuting operators $\{ \hat P^i\}$
of NP whose eigenvalues $p^i$ are components of the momentum ${\bf p}$. Among
different versions of solutions which are currently used we prefer the
normalized solutions of positive frequencies that read
\begin{equation}
f_{\bf
p}(x)=\frac{1}{2}\sqrt{\frac{\pi}{\omega}}\frac{1}{(2\pi)^{3/2}}\,e^{-3\omega
t/2}Z_k\left(\frac{p}{\omega}\,e^{-\omega t}\right) e^{i {\bf p}\cdot {\bf
x}}\,,
\end{equation}
where the functions ${Z}_k$ are defined in the Appendix A, $p=|{\bf p}|$ and we
denote
\begin{equation}\label{k}
k=\sqrt{\mu^2-\textstyle{\frac{9}{4}}}\,, \quad \mu=\frac{m}{\omega}\,,
\end{equation}
provided $m>3\omega/2$. Obviously, the fundamental solutions of negative
frequencies are $f_{\bf p}^*(x)$. All these solutions satisfy the
orthonormalization relations
\begin{eqnarray}
\langle  f_{\bf p},f_{{\bf p}'}\rangle=-\langle  f_{\bf p}^*,f_{{\bf
p}'}^*\rangle&=&\delta^3({\bf p}-{\bf p}')\,,\\
\langle f_{\bf p},f_{{\bf p}'}^*\rangle&=&0\,,
\end{eqnarray}
and the completeness condition
\begin{equation}\label{comp}
i\int d^3p\,  f^*_{\bf p}(t,{\bf x}) \stackrel{\leftrightarrow}{\partial_{t}}
f_{\bf p}(t,{\bf x}')=e^{-3\omega t}\delta^3({\bf x}-{\bf x}')\,.
\end{equation}
For this reason  we say that the set $\{ f_{\bf p}| {\bf p}\in {\Bbb R}^3_p\}$
forms the complete system of fundamental solutions of the {\em momentum} basis
of NP. In this basis, the KG field can expanded in terms of plane waves of
positive and negative frequencies in usual manner as
\begin{equation}\label{field1}
\phi(x)=\phi^{(+)}(x)+\phi^{(-)}(x)=\int d^3p \left[f_{\bf p}(x)a({\bf
p})+f_{\bf p}^*(x)b^*({\bf p})\right]
\end{equation}
where $a$ and $b$ are the wave functions of the momentum representation that
can be calculated using the inversion formulas
\begin{equation}
a({\bf p})=\langle f_{\bf p},\phi\rangle\,,\quad b({\bf p})=\langle f_{\bf
p},\phi^*\rangle\,.
\end{equation}

\subsection{The energy basis}

The plane waves of given energy have to be derived in the SP where the KG
equation has the suitable form (\ref{KG2}). We assume that in this picture the
KG field can be expanded as
\begin{eqnarray}
&&\phi_S(x)=\phi_S^{(+)}(x)+\phi_S^{(-)}(x)\nonumber\\
&& =\int_{0}^{\infty} dE \int d^3q \left[\hat\phi_S^{(+)}(E,{\bf
q})e^{-i(Et-{\bf q}\cdot{\bf x})}+ \hat\phi_S^{(-)}(E,{\bf q})e^{i(Et-{\bf
q}\cdot{\bf x})}\right]\label{KGS}
\end{eqnarray}
where $\hat{\phi}^{(\pm)}_S$  behave as tempered distributions on the domain
${\Bbb R}_q^3$ such that the Green theorem may be used. Then we can replace the
momentum operators ${P}_S^i$ by  ${q}^i$ and the coordinate operators ${X}_S^i$
by $i {\partial}_{q_i}$ obtaining the KG equation of the SP in momentum
representation,
\begin{eqnarray}
&&\left\{\left[\pm i E+\omega \left({q}^i{
\partial}_{q_i}+3\right)\right]^2\right.\nonumber\\
&&~~~~~\left. -\, 3\omega \left[\pm i E+\omega \left({q}^i{
\partial}_{q_i}+3\right)\right]+{\bf q}^2+m^2\right\}
\hat{\phi}^{(\pm)}_S(E,{\bf q})=0\,,\label{KG4}
\end{eqnarray}
where $E$ is the energy defined as the eigenvalue of $H_S$. We remind the
reader that the operators ${P}^i_S$ and ${X}^i_S$ become in NP the time
dependent operators (\ref{Xt}) and respectively (\ref{Pt}) while $H_S$ is
related through Eq. (\ref{Ht}) to the conserved energy operator $H$.  This
means the energy $E$ is a conserved quantity but the momentum ${\bf q}$ does
not have this property. More specific, only the scalar momentum $q=|{\bf q}|$
is not conserved while the momentum direction is conserved since the operator
(\ref{Pt}) is parallel  with the conserved momentum $\hat{\bf P}$.   For this
reason we denote ${\bf q}=q\, {\bf n}$ observing that the differential operator
of Eq. (\ref{KG4}) is of radial type and reads
${q}^i{\partial}_{q_i}=q\,\partial_q$. Consequently, this operator acts only on
the functions depending on $q$ while the functions which depend on the momentum
direction ${\bf n}$ behave as constants. Therefore, we have to look for
solutions of the form
\begin{eqnarray}
\hat{\phi}^{(+)}_S(E,{\bf q})&=&h_S(E,{q})\,a(E,{\bf n})\,,\\
\hat{\phi}^{(-)}_S(E,{\bf q})&=&[h_S(E,{q})]^*\,b^*(E,{\bf n})\,,
\end{eqnarray}
where the function $h_S$ satisfies an equation derived from Eq. (\ref{KG4})
that can be written simply in the new variable $s=q/\omega$ and using the
notations (\ref{k}) and $\epsilon=E/\omega$. This equation,
\begin{equation}
\left[\frac{d^2}{ds^2}+\frac{2i\epsilon+4}{s}\frac{d}{ds}
+\frac{\mu^2-\epsilon^2+3i\epsilon}{s^2}+1\right] h_S(\epsilon, s)=0\,,
\end{equation}
is of the Bessel type having solutions of the form $h_S(\epsilon,s)={\rm
const}\,\, s^{-i\epsilon -3/2}\,Z_k(s)$. Collecting all the above results we
derive the final expression of the KG field (\ref{KGS}) as
\begin{equation}\label{KGSSP}
\phi_S(x)=\int_0^{\infty}\,dE\int_{S^2}\, d\Omega_n\,\, \left\{f^S_{E,{\bf
n}}(x) a(E,{\bf n}) + \, [f^S_{E,{\bf n}}(x)]^* b^*(E,{\bf n})\right\}\,,
\end{equation}
where the integration covers the sphere $S^2\subset {\Bbb R}^3_p$. The
fundamental solutions $f^S_{E,{\bf n}}$ of positive frequencies, with energy
$E$ and momentum direction ${\bf n}$ result to have the integral representation
\begin{equation}\label{fps}
f^S_{E,{\bf n}}(x)=
 Ne^{-iEt}\int_{0}^{\infty} ds\,\sqrt{s}\,\, Z_k(s)\,
 e^{i \omega s {\bf n}\cdot{\bf x}-i\epsilon \ln s}\,,
\end{equation}
where $N$ is a normalization constant.

For understanding the physical meaning of this result we must turn back to NP
where the scalar field
\begin{equation}\label{field2}
\phi(x)=\int_0^{\infty}\,dE\int_{S^2}\, d\Omega_n\,\, \left\{f_{E,{\bf n}}({
x}) a(E,{\bf n}) + \, [f_{E,{\bf n}}({x})]^* b^*(E,{\bf n})\right\}\,,
\end{equation}
is expressed in terms of the solutions of NP which can be calculated as
\begin{equation}\label{fpsNP}
f_{E,{\bf n}}({x})=U^{-1}(x)f^S_{E,{\bf n}}({x})=
 Ne^{-iEt}\int_{0}^{\infty} ds\,\sqrt{s}\, \, Z_k(s)\,
  e^{i \omega s {\bf n}\cdot{\bf x}_t-i\epsilon \ln
 s}\,,
\end{equation}
where ${\bf x}_t=e^{\omega t}{\bf x}$. Finally, changing the integration
variable, $e^{\omega t}s\to s$, we obtain the definitive integral
representation
\begin{equation}\label{fdef}
f_{E,{\bf n}}({x})=
 Ne^{-3\omega t/2}\int_{0}^{\infty} ds\, \sqrt{s}\, \, Z_k\left(s e^{-\omega t}\right)\,
  e^{i \omega s {\bf n}\cdot{\bf x}-i\epsilon \ln
 s}\,.
\end{equation}
Now using the scalar product (\ref{SP2}) and the method of the Appendix B we
can show that the normalization constant
\begin{equation}\label{norm}
N=\frac{1}{2} \sqrt{\frac{\omega}{2}}\frac{1}{(2\pi)^{3/2}}\,,
\end{equation}
assures the desired orthonormalization relations
\begin{eqnarray}
\langle f_{E,{\bf n}},f_{E',{\bf n}^{\,\prime}}\rangle=- \langle f^*_{E,{\bf
n}},f^*_{E',{\bf n}^{\,\prime}}\rangle&=& \delta(E-E')\,\delta^2 ({\bf n}-{\bf
n}^{\,\prime})\,,
\label{orto1}\\
\langle f_{E,{\bf n}},f^*_{E',{\bf n}^{\,\prime}}\rangle&=&0\,, \label{orto2}
\end{eqnarray}
and the completeness condition
\begin{equation}\label{comp1}
i\int_0^{\infty}dE\int_{S^2} d\Omega_n \left\{ [f_{E,{\bf n}}(t,{\bf
x})]^*\stackrel{\leftrightarrow}{\partial_{t}} f_{E,{\bf n}}(t,{\bf x}')
\right\} =e^{-3\omega t}\delta^3 ({\bf x}-{\bf x}^{\,\prime})\,.
\end{equation}
This means that the set of functions $\{f_{E,{\bf n}}|E\in{\Bbb R}^+, {\bf
n}\in S^2\}$ constitutes the complete system of fundamental solutions of the
{\em energy} basis of the NP.

The last step is to calculate the transition coefficients between the momentum
and energy bases of the NP that read
\begin{equation}
\langle f_{\bf p},f_{E,{\bf n}}\rangle=\langle f_{E,{\bf n}},f_{\bf
p}\rangle^*=\frac{p^{-3/2}}{\sqrt{2\pi\omega}}\,\delta^2({\bf n}-{\bf
n}_p)\,e^{-i\frac{E}{\omega}\ln \frac{p}{\omega}}\,,
\end{equation}
where ${\bf n}_p={\bf p}/p$. With their help we deduce the transformations
\begin{eqnarray}
a({\bf p})&=&\int_0^{\infty}dE\int_{S^2}d\Omega_n \langle f_{\bf p},f_{E,{\bf
n}}\rangle a(E,{\bf n})\nonumber\\
&=&\frac{p^{-3/2}}{\sqrt{2\pi\omega}}\int_0^{\infty}dE\,
e^{-i\frac{E}{\omega}\ln \frac{p}{\omega}}\,a(E,{\bf n}_p)\,,\label{Iaa1}\\
a(E,{\bf n})&=&\int d^3 p \,\langle f_{E,{\bf n}},f_{\bf p}\rangle a({\bf
p})\nonumber\\
&=&\frac{1}{\sqrt{2\pi\omega}}\int_0^{\infty}dp\,\sqrt{p}\,\,
e^{\,i\frac{E}{\omega}\ln \frac{p}{\omega}}\,a(p\,{\bf n})\,,\label{Iaa2}
\end{eqnarray}
and similarly for the wave functions $b$. These relations  will help us to
perform the second quantization in canonical manner using  both the bases
defined above.

\section{Quantization}

The quantization can be done in canonical manner considering that the wave
functions $a$ and $b$ of the fields (\ref{field1}) and (\ref{field2}) become
field operators (such that $b^{*}\to b^{\dagger}$) \cite{SW1}. We assume that
the particle ($a$, $a^{\dagger}$) and antiparticle ($b$, $b^{\dagger}$)
operators fulfill the standard commutation relations in the momentum basis,
from which the non-vanishing ones are
\begin{equation}\label{com1}
[a({\bf p}), a^{\dagger}({\bf p}^{\,\prime})]=[b({\bf p}), b^{\dagger}({\bf
p}^{\,\prime})] = \delta^3 ({\bf p}-{\bf p}^{\,\prime})\,.
\end{equation}
Then, from Eq. (\ref{Iaa1}) it results that the field operators of the energy
basis satisfy
\begin{equation}\label{com2}
[a(E,{\bf n}), a^{\dagger}(E',{\bf n}^{\,\prime})]=[b(E,{\bf n}),
b^{\dagger}(E',{\bf n}^{\,\prime})] = \delta(E-E') \delta^2 ({\bf n}-{\bf
n}^{\,\prime})\,,
\end{equation}
and
\begin{equation}
[a({\bf p}), a^{\dagger}(E,{\bf n})]=\langle f_{\bf p},f_{E,{\bf n}}\rangle\,,
\end{equation}
while other commutators are vanishing.  In this way  the field $\phi$ is
correctly quantized according to the {\em canonical} rule
\begin{equation}
[ \phi(t,{\bf x}),\pi(t,{\bf x}')]=e^{3\omega t}\,[ \phi(t,{\bf
x}),\partial_{t}\phi^{\dagger}(t,{\bf x}')]=i\,\delta^3({\bf x}-{\bf x}')\,,
\end{equation}
where $\pi=\sqrt{g}\,\partial_t\phi^{\dagger}$ is the momentum density derived
from the action (\ref{action}). All these operators act on the Fock space
supposed to have an unique vacuum state $|0\rangle$ accomplishing
\begin{equation}
a({\bf p})|0\rangle=b({\bf p})|0\rangle=0\,,\quad \langle0|a^{\dagger}({\bf
p})=\langle 0|  b^{\dagger}({\bf p})=0\,,
\end{equation}
and similarly for the energy basis. The sectors with a given number of
particles have to be constructed using the standard methods, obtaining thus the
generalized bases of momentum or energy.

Furthermore, we have to calculate the one-particle operators corresponding to
the conserved quantities (\ref{cLAB}). This can be achieved bearing in mind
that for any self-adjoint generator $A$ of the scalar representation of the
group $I(M)$ there exists a {\em conserved} one-particle operator of the
quantum field theory which can be calculated simply as
\begin{equation}\label{opo}
{\cal A}=:\langle \phi, A\phi\rangle:
\end{equation}
respecting the normal ordering of the operator products \cite{SW1}. Hereby we
recover the standard algebraic properties
\begin{equation}\label{algXX}
[{\cal A}, \phi(x)]=-A\phi(x)\,, \quad [{\cal A}, {\cal B}\,]=:\langle \phi,
[A,B]\,\phi\rangle:
\end{equation}
due to the canonical quantization adopted here. In other respects, the electric
charge operator corresponding to the $U(1)$ internal symmetry  (of Abelian
gauge transformations $\phi\to e^{i\alpha I} \phi)$ results from the Noether
theorem to be ${\cal Q}=:\langle \phi, I\phi\rangle:={:\langle \phi,
\phi\rangle:}$.

However, there are many other conserved operators which do not have
corresponding differential operators at the level of quantum mechanics.  The
simplest examples are the operators of number of particles,
\begin{equation}
{\cal N}_{pa}=\int d^3p\,   a^{\dagger}({\bf p}) a({\bf p})=\int_0^{\infty}dE
\int_{S^2} d\Omega_n   a^{\dagger}(E,{\bf n}) a(E,{\bf n})\,,
\end{equation}
and that of antiparticles, ${\cal N}_{ap}$ (depending on $b$ and
$b^{\dagger}$), which give the charge operator ${\cal Q}={\cal N}_{pa}-{\cal
N}_{ap}$ and the operator of total number of particles, ${\cal N}={\cal
N}_{pa}+{\cal N}_{ap}$.

In what follows we focus on the conserved one-particle operators determining
the momentum and energy bases. The diagonal operators of the momentum basis the
are ${\cal Q}$ and the components of momentum operator,
\begin{equation}
{\cal P}^i=:\langle \phi, \hat P^i\phi\rangle:=\int d^3p\, p^i \left[
a^{\dagger}({\bf p}) a({\bf p})+b^{\dagger}({\bf p}) b({\bf p})\right]\,.
\end{equation}
In other words, the momentum basis is determined by the set of commuting
operators $\{ {\cal Q}, {\cal P}^i\}$. The energy basis is formed by the common
eigenvectors of the set of commuting operators $\{ {\cal Q}, {\cal H},
\tilde{\cal P}^i\}$, i.e. the charge, energy and momentum direction operators.
The energy operator can be easily calculated since the solutions (\ref{fpsNP})
are eigenfunctions of the operator $H$. In this way we find
\begin{equation}
{\cal H}=:\langle \phi, H\phi\rangle:=\int_0^{\infty} dE\, E \int_{S^2}
d\Omega_n \left[ a^{\dagger}(E,{\bf n}) a(E,{\bf n})+b^{\dagger}(E,{\bf n})
b(E,{\bf n})\right]\,.
\end{equation}
More interesting are the operators $\tilde{\cal P}^i$ of the momentum direction
since they do not come from differential operators and, therefore, must be
defined directly as
\begin{equation}
\tilde{\cal P}^i=\int_0^{\infty} dE \int_{S^2} d\Omega_n \, n^i\,\left[
a^{\dagger}(E,{\bf n}) a(E,{\bf n})+b^{\dagger}(E,{\bf n}) b(E,{\bf
n})\right]\,.
\end{equation}
The above operators which satisfy simple commutation relations,
\begin{equation}\label{comHP}
[{\cal H}, {\cal P}^i]=i\omega {\cal P}^i\,,\quad [{\cal H},\tilde{\cal
P}^i]=0\,,\quad [{\cal Q}, {\cal H}]=[{\cal Q}, {\cal P}^i]=[{\cal Q},\tilde
{\cal P}^i]=0\,,
\end{equation}
are enough for defining the bases considered hare. However, there are more six
conserved operators corresponding to the differential operators $L_i$ and
$R^i$, defined by Eqs. (\ref{LLL1}) and respectively (\ref{LLL2}), but their
calculation according to the rule (\ref{opo}) requires special methods which
will be discussed elsewhere.

Our approach offers the opportunity to provide closed expressions for the
conserved one-particle operators in bases where these are not diagonal. For
example we can calculate the energy operator in momentum basis either starting
with the identity
\begin{equation}
(Hf_{\bf p})(x)=-i\omega \left(p^i\partial_{p_i}+{\frac{3}{2}}\right)f_{\bf
p}(x)
\end{equation}
or using Eq. (\ref{Iaa2}). The final result,
\begin{equation}\label{Hpp}
{\cal H}=\frac{i\omega}{2}\int d^3p\, p^i \left\{ \left[\, a^{\dagger}({\bf
p})\stackrel{\leftrightarrow}{\partial}_{p_i} a({\bf p})\right]+ \left[\,
b^{\dagger}({\bf p}) \stackrel{\leftrightarrow}{\partial}_{p_i} b({\bf
p})\right]\right\}\,,
\end{equation}
is similar with that obtained in Ref. \cite{CD1} for the Dirac field on $M$.

We note that beside the above conserved operators we can introduce other
one-particle operators extending the definition (\ref{opo}) to the
non-conserved operators of our quantum mechanics. However, these operators will
depend explicitly on time, their expressions being complicated and without an
intuitive physical meaning.

In the quantum theory of fields it is important to study  the Green functions
related to the partial commutator functions (of positive or negative
frequencies) defined as
\begin{equation}
D^{(\pm)}(x,x')= i[\phi^{(\pm)}(x),\phi^{(\pm)\,\dagger}(x')]
\end{equation}
and the total one, $D=D^{(+)}+D^{(-)}$. These function are solutions of the KG
equation in both the sets of variables and obey
$[D^{(\pm)}(x,x')]^*=D^{(\mp)}(x,x')$  such that $D$ results to be a real
function. This property suggest us to restrict ourselves to study only the
functions of positive frequencies,
\begin{equation}
D^{(+)}(x,x')=i\int d^3 p \, f_{\bf p}(x)f_{\bf p}(x')^* =i\int_0^{\infty} dE
\int_{S^2} d\Omega_n \, f_{E,\bf n}(x)f_{E,\bf n}(x')^*\,,
\end{equation}
resulted from  Eqs. (\ref{field1}) and (\ref{field2}). Both these versions lead
to the final expression
\begin{eqnarray}
D^{(+)}(x,x')&=&\frac{\pi}{4\omega}\frac{i}{(2\pi)^3}\,
e^{-\frac{3}{2}\,\omega(t+t')}\nonumber\\
&&\times \int d^3 p\, Z_k\left(\frac{p}{\omega}\,e^{-\omega
t}\right)Z_k^*\left(\frac{p}{\omega}\,e^{-\omega t'}\right)e^{i{\bf
p}\cdot({\bf x}-{\bf x}')}
\end{eqnarray}
from which we understand that $D^{(+)}(x,x')=D^{(+)}(t,t',{\bf x}-{\bf x}')$
and may deduce what happens at equal time. First we observe that for $t'=t$ the
values of the function $D^{(+)}(t,t,{\bf x}-{\bf x}')$ are c-numbers which
means that $D(t,t,{\bf x}-{\bf x}')=0$. Moreover, from Eqs. (\ref{comp}) or
(\ref{comp1}) we find
\begin{equation}\label{Dtt}
\left.(\partial_t -\partial_{t'})D^{(+)}(t,t',{\bf x}-{\bf
x}')\right|_{t'=t}=e^{-3\omega t}\delta^3({\bf x}-{\bf x}')
\end{equation}
and similarly for $D^{(-)}$.

These functions help us to introduce the principal Green functions.  In
general, $G(x,x')=G(t,t',{\bf x}-{\bf x}')$ is a Green function of the KG
equation if this obeys
\begin{equation}\label{KGG}
\left( \partial_t^2-e^{-2\omega t}\Delta_x +3\omega
\partial_t+m^2\right)G(x,x')=e^{-3\omega t}\delta^4(x-x')\,.
\end{equation}
The properties of the commutator functions allow us to construct the Green
function just as in the scalar theory on Minkowski spacetime. We assume that
the retarded, $D_R$, and advanced, $D_A$, Green functions read
\begin{eqnarray}
D_R(t,t',{\bf x}-{\bf x}')&=& \theta(t-t')D(t,t',{\bf x}-{\bf x}')\,,\\
D_A(t,t',{\bf x}-{\bf x}')&=& -\,\theta(t'-t)D(t,t',{\bf x}-{\bf x}')\,,
\end{eqnarray}
while the Feynman propagator,
\begin{eqnarray}
&&D_F(t,t',{\bf x}-{\bf x}')= i\langle 0|T[\phi(x)\phi^{\dagger}(x')]\,|0\rangle\nonumber\\
&&= \theta (t-t') D^{(+)}(t,t',{\bf x}-{\bf x}')-\theta(t'-t)D^{(-)}(t,t',{\bf
x}-{\bf x}')\,,
\end{eqnarray}
is defined as a causal Green function. It is not difficult to verify that all
these functions satisfy Eq. (\ref{KGG}) if one uses the identity
$\partial_t^2[\theta(t)f(t)]=\delta(t)\partial_t f(t)$, the artifice
$\partial_t f(t-t')=\frac{1}{2}(\partial_t-\partial_{t'}) f(t-t')$ and Eq.
(\ref{Dtt}).

These Green functions have complicated structures such that it would be helpful
to analyze their properties in momentum representation. However, this can not
be achieved in NP where Eq. (\ref{KGG}) does not have a convenient Fourier
transformation because of its time dependence. Thus, the real challenge in
further investigations will be the study of the Green functions in SP where
this equation can be solved in momentum representation.

\section{Concluding remarks}

We presented here the complete quantum theory of the massive charged scalar
field minimally coupled with the gravitation of the dS  expanding universe,
$M$. The main point of our approach is the new set of fundamental solutions of
the energy basis, derived with the help of our SP. This new basis completes the
framework of quantum theory, being crucial for understanding how can be
measured the energy and momentum whose operators do not commute among
themselves.

In our opinion there exists a {\em global} apparatus providing quantum
observables globally defined on $M$, which have different forms in each local
chart but global algebraic properties which do not depend on the chart we
choose. In our case the global apparatus offer us a large collection of
conserved operators corresponding to the dS isometries or some internal
symmetries, determining quantum modes whose wave functions are: (I) solutions
of the KG equation and (II) eigenfunctions of suitable systems of commuting
conserved operators. In other words, the quantum states of the scalar field on
$M$ can be {\em prepared} only by the global apparatus since the KG equation is
global.

Let us see how the global apparatus  measures the energy and momentum of the
one-particle state defined as
\begin{equation}
|\chi\rangle=\int d^3 p\, \chi({\bf p})a^{\dagger}({\bf p})|0\rangle\,,\quad
\chi({\bf p})=\rho({\bf p})e^{-i\vartheta({\bf p})}\,,
\end{equation}
where $\rho$ and $\vartheta$ are real functions. The normalization condition
\begin{equation}\label{normro}
1=\langle\chi|\chi\rangle= \int d^3 p\, |\chi({\bf p})|^2=\int d^3 p\,
|\rho({\bf p})|^2
\end{equation}
shows that the function $\rho$ must be square integrable on ${\Bbb R}^3_p$
while $\vartheta$ remains an arbitrary function. Calculating the expectation
values of the non-commuting operators ${\cal P}^i$ and ${\cal H}$,  we obtain
first the usual formula of the momentum expectation value
\begin{equation}
\langle\chi|{\cal P}^i|\chi\rangle= \int d^3 p\,p^i\, |\rho({\bf p})|^2\,.
\end{equation}
which is independent on $\vartheta$. However, for the energy operator the
result is rather surprising since from Eq. (\ref{Hpp}) we derive an expectation
value
\begin{equation}
\langle\chi|{\cal H}|\chi\rangle=\omega  \int d^3 p\,[p^i\partial_{p_i}
\vartheta({\bf p})]\,|\rho({\bf p})|^2\
\end{equation}
which depends mainly on the phase $\vartheta$. This means that  we can prepare
at anytime states with arbitrary desired expectation values of these
observables. Thus, in the particular case of $\vartheta ({\bf p})=\epsilon
\ln(p)$ we obtain $\langle\chi|{\cal H}|\chi\rangle=\omega\epsilon=E$
indifferent on the form of $\rho$ if this obeys the condition (\ref{normro}).

Beside the global apparatus, one may use other types of {\em local} apparatus
\cite{BD} measuring quantum modes whose wave functions do not satisfy the
condition (I). These form bases of the Fock space determined by sets of
commuting operators which do not commute with the KG one. Consequently, the
local apparatus can measure the parameters of some arbitrary quantum states in
a given state prepared by the global apparatus but it is not able to prepare
itself states of the KG field, remaining a mere detector. Moreover, when one
detects quantum modes that are not genuine KG ones and, therefore, can not be
correctly normalized, it appears the danger of creating  the welling illusion
even on $M$ where the vacuum is stable.

We conclude that our approach seems to be coherent at the level of relativistic
quantum mechanics where we give correct definitions of the principal
observables and introduce the SP allowing us to find the energy basis. For this
reason the second quantization in canonical manner leads to quantum free fields
which can be manipulated simply as those of special relativity. These fields
may be used now for calculating scattering processes on the dS expanding
universe.

\appendix

\subsection*{Appendix A: Special Bessel functions}

Let us consider the Hankel functions $H^{(1,2)}_{\nu}(s)$ in the special case
when $\nu=i k$ and define the functions $Z_k$ such that
\begin{equation}
Z_k(s)=e^{-\pi k/2}H^{(1)}_{ik}(s)\,,\quad Z_k^*(s)=e^{\pi
k/2}H^{(2)}_{ik}(s)\,.
\end{equation}
Then,  using the Wronskian $W$ of the Bessel functions \cite{AS} we find that
\begin{equation}\label{ZuZu}
Z_k^*(s)
\stackrel{\leftrightarrow}{\partial_{s}}Z_k(s)=W[H_{ik}^{(2)},H_{ik}^{(1)}]=\frac{4i}{\pi
s}\,.
\end{equation}

\subsection*{Appendix B: Normalization integrals}

In spherical coordinates of the momentum space, ${\bf n}\sim
(\theta_n,\phi_n)$, and the notation ${\bf q}=\omega s{\bf n}$, we have $d^3q=
q^2dq\, d\Omega_n=\omega^3\, s^2ds\, d\Omega_n$ with
$d\Omega_n=d(\cos\theta_n)d\phi_n$. Moreover, we can write
\begin{equation}\label{del}
\delta^3({\bf q}-{\bf q}^{\,\prime})=\frac{1}{q^2}\,\delta(q-q')\delta^2({\bf
n}-{\bf n}') =\frac{1}{\omega^3 s^2}\,\delta(s-s')\delta^2({\bf n}-{\bf n}')\,,
\end{equation}
where we denoted $\delta^2({\bf n}-{\bf n}') =\delta(\cos \theta_n-\cos
\theta_n')\delta(\phi_n-\phi_n')\,.$

The normalization integrals can be calculated either in SP using the scalar
product (\ref{spSP}) which leads to complicated calculations or simply in NP
starting with the scalar product (\ref{SP2}). According to Eqs. (\ref{fdef})
and (\ref{del}), this yields
\begin{eqnarray}
\langle f_{E,{\bf n}},f_{E',{\bf n}^{\,\prime}}\rangle&=&i\,\frac{N^2
(2\pi)^3}{\omega^3}\, \delta^2({\bf n}-{\bf n}')\int_0^{\infty}
\frac{ds}{s}e^{i(\epsilon-\epsilon')\ln s}\nonumber\\
&&\hspace*{12mm}\times \left[Z_k^*(s\,e^{-\omega
t})\stackrel{\leftrightarrow}{\partial_{t}}Z_k(s\,e^{-\omega t})\right]\,.
\end{eqnarray}
Furthermore, using Eq. (\ref{ZuZu}) and the identity
\begin{equation}
\frac{1}{2\pi\omega}\int_0^{\infty}\frac{ds}{s}\, e^{i(\epsilon-\epsilon')\ln
s} =\delta(E-E')\,,
\end{equation}
we find the value of the normalization constant (\ref{norm}).

\end{document}